# Development of age spots as a result of accumulation of aged cells in aged skin

Jicun Wang-Michelitsch[1]*, Thomas M Michelitsch[2]

[1]Department of Medicine, Addenbrooke's Hospital, University of Cambridge, UK (work address until 2007)

[2]Institut Jean le Rond d'Alembert (Paris 6), CNRS UMR 7190 Paris, France

## Abstract

Age spots are the brown spots that develop in the skin but change in color and shape with time. To understand the mechanism of development of age spots, characteristics of age spots are analyzed by Misrepair mechanism, a mechanism introduced in Misrepair-accumulation aging theory. An age spot is pathologically a group of aggregated basal cells, which contain lipofuscin bodies. Accumulation of lipofuscin bodies is a sign of aging of a cell. Characteristics of age spots include: in-homogeneity in distribution, growing flatly before becoming protruding, irregularity on shape, in-homogeneity on the color and on the protruding degree of a spot, and softness of a protruding spot. After analyzing these characteristics, we make a hypothesis on the process of development of an age spot. **A**. Aging of a tissue is the basis for development of age spots. Deposition of a lipofuscin-containing cell is a consequence of aging of the skin, and it is a manifestation of Misrepair. **B**. A flat spot results from accumulation of lipofuscin-containing cells. When an aged cell remains, this cell can accelerate the aging of its neighbor cells by increasing damage-sensitivity and reducing repair-efficiency of the local tissue. By a viscous circle, more and more neighbor cells become aged and they form a flat spot, which has an irregular shape. **C**. A protruding spot develops when some of the cells in a flat spot die and release lipofuscin bodies. For the survival of an organism, the un-degradable lipofuscin bodies have to be isolated by a capsule made by fibrotic membrane, for maintaining the structural integrity of local epidermis. Successive deaths of lipofuscin-containing cells make the capsule include more and more dead substances by layers of fibrotic membrane. In this way, the spot "grows" in three-dimension, resulting in protruding of the spot. In conclusion, development of an age spot is a result of accumulation of aged cells in aged skin.

**Introduction**

Age spots are the brown spots on the skin, but they are different from melanotic nevi and verrucous nevi. Melanotic nevi are the spots that are flat or mildly protruding with homogenous color: deep brown or black. Melanotic nevi are all round in shape with clear boundary. Melanotic nevi can develop at any age, and they are pathologically a group of melanocytes, which can be a result of cell proliferation. A single verrucous epidermal nevus is a spot that is round and protruding with color of light brown, and it can develop also at any age. Differently, age spots are the spots that are in the beginning flat with light color and irregular shape but later become larger, protruding, and darker in color. An age spot is pathologically a group of basal cells, which contain lipofuscin bodies. Lipofuscin bodies are intracellular lysosomes, which contain half-digested cell wastes. Age spots develop also in the liver, cordial wall, and nerve system; therefore they are also called "lipofuscin spots" and "liver spots". Since accumulation of lipofuscin bodies in a cell is a sign of aging of the cell, age spots is a result of aggregation of aged basal cells. So far, it is unknown why these aged cells cluster together and why the spots are changing with time on size, shape, and color. To these phenomena, traditional aging theories including damage (fault)-accumulation theory [1] and gene-controlling theory [2, 3] are unfortunately unable to give an interpretation. In the present paper, we will analyze age spots with Misrepair mechanism, and make a hypothesis on the process of development of an age spot. Misrepair mechanism is a mechanism that we have proposed in Misrepair-accumulation aging theory [4, 5]. Our discussion in this paper tackles the following issues:

I.    Characteristics of age spots

II.   A novel aging theory: Misrepair-accumulation theory

III.  Misrepair mechanism in the development of an age spot

   3.1  Deposition of lipofuscin-containing cells in a tissue

   3.2  Enlargement of a flat spot as a result of accumulation of aged cells

   3.3  Protruding of a spot as a result of isolation of lipofuscin bodies in a tissue

IV.   Conclusions

**I.    Characteristics of age spots**

Age spots develop mainly on the part of skin which is often exposed to sunlight, including the face and the back sides of hands. The brown color of an age spot is due to the lipofuscin bodies in cells. Lipofuscin is a mixture of lipids and proteins in lysosomes, in which lipids bind to protein fragments via malondialdehyde. Lipofuscin inclusion bodies are lysosomes that contain half-digested cell wastes, and they accumulate when a cell has reduced efficiency on functions or has increased production of wastes. Thus accumulation of lipofuscin bodies is a result but not the cause of aging of a cell. Lipofuscin bodies have been also seen in neuron cells, muscle cells, and hepatocytes. Within part of the skin, age spots are different from each other on shape, size, color, and the degree of protruding (Figure 1). New spots tend to develop close to old ones, resulting in a satellite-like distribution of spots. Old spots are permanently larger than new ones. A flat spot is light brown in color and is smoothing like that in normal skin. A protruding spot is a soft spot, dark in color, covered with cuticle-like substance, and filled with lipids. In a protruding spot, basal cells aggregate in columns from the level of basement membrane to skin surface, accompanied by thickening of squamous epithelial layers and keratinocyte layers.

During the development of age spots, some phenomena are characteristic. **A**. Age spots are inhomogeneous on distribution, size, and shape. **B**. A flat spot can "grow" in two-dimension and rest "flat" for many years before becoming protruding. **C**. Within one spot, some parts can be flat and some parts are protruding with deeper color, which makes a spot rough and in-homogenous in color**. D.** A protruding spot in late stage is soft, lipid-containing, but irremovable (Figure 1). A sound aging theory should be able to explain all of these phenomena. However, for the in-homogeneous distribution of spots, none of traditional theories can give an interpretation. For example, gene-controlling theory suggests that certain genes control the whole process of aging independently [2, 3]. However, if such genes existed, the locations of age spots should be independent of the locations of damage-exposure, which is not true. Damage (fault)-accumulation theory suggests that aging is a result of accumulation of "faults" as intrinsic damage, which are left unrepaired due to limitation of repair/maintenance [1]. However, if aging was a direct result of damage, accumulation of random damage should result in a homogenous distribution of spots within the affected area of tissue. For understanding the development of age spots, a distinct view on aging is needed. We will show that Misrepair-accumulation theory can explain the in-homogeneity in the development of age spots [4, 5].

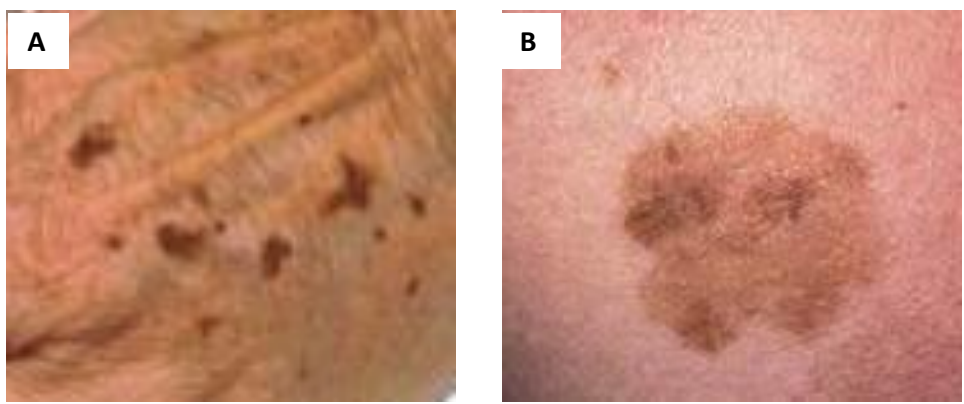

**Figure 1. In-homogeneity of age spots on distribution, on size, and on shape**

The age spots on the skin are inhomogeneous on distribution, on size, and on shape **(A)**. Even within one spot, some parts can be flat with light color and some parts are protruding with deeper color, which makes the spot rough and inhomogeneous in color (**B**).

## II.    A novel aging theory: Misrepair-accumulation theory

For explaining aging changes, we have proposed a generalized concept of Misrepair in our Misrepair-accumulation theory [4, 5]. The new concept of Misrepair is defined as ***an incorrect reconstruction of an injured living structure,*** and it is applicable to all living structures including molecules (DNAs), cells, and tissues. Scar formation is a typical example of Misrepair. In situations of severe injuries, when a complete repair is impossible to achieve, Misrepair is a strategy of repair for maintaining the structural integrity and for the survival of an organism. However, Misrepair results in alteration of the structure and reduction of the functionality of a living structure. The structural alterations made by Misrepairs are irreversible and irremovable, and they accumulate and deform gradually a living structure, leading to aging of it. Thus, aging of an organism is a process of accumulation of Misrepairs. Misrepair mechanism is a surviving mechanism for an organism; and it is essential for the survival of a species.  Aging of an individual is a sacrifice for species' survival.

Misrepairs have a tendency to accumulate to the part of a tissue where an old Misrepair has taken place, since this part of tissue has increased damage-sensitivity and reduced repair-efficiency. Accumulation of Misrepairs is therefore focalized and self-accelerating. Development of aging changes is thus self-accelerating and inhomogeneous [6]. Aging can take place on each level of living structures; however aging of an organism takes place essentially on tissue level. An irreversible change of the spatial relationship between cells and/or extracellular matrixes in a tissue is ***essential and sufficient*** for causing a decline of organ functionality. Aging of a tissue does not always require aging of cells. In contrast, aging of a tissue is often the cause for aging of cells.

## III.    Misrepair mechanism in the development of an age spot

An age spot is pathologically a group of aggregated basal cells, which contain lipofuscin bodies. It is so far unknown whether or not this cell-aggregation is a result of cell proliferation. In our view, a flat spot cannot be a result of cell proliferation by four reasons. Firstly, if the cells in a spot are produced by proliferation of a lipofuscin-containing cell, these cells should have reduced levels of lipofuscin and the spot should have reduced color during growing, which is however not true. Secondly, a flat spot can grow flatly for many years, and this may not be a result of cell proliferation. Clonal expansion of cells by cell proliferation undergoes more or less in three-dimension in a tissue. Thirdly, if development of a spot is a result of cell proliferation, the boundary of the spot should be smoothing rather than irregular. Finally, lipofuscin-containing cells are dying cells with reduced potential of cell division, thus they cannot proliferate for many years. Therefore, the aggregation of cells in an aged spot is more likely a result of accumulation of aged cells. But the questions are: why these basal cells in part of a tissue undergo aging successively; and what the factor is that drives the process of successive aging of cells in a neighborhood. We will try to answer these two questions in this part by Misrepair mechanism.

## 3.1    Deposition of lipofuscin-containing cells in an aged tissue

Lipofuscin bodies are cell wastes, which are half-degraded and isolated in lysosomes in a cell. Appearance of lipofuscin bodies is a sign of aging of a cell. For the cells with longer lifespan and longer process of aging such as neuron cells and muscle cells, intracellular lipofuscin bodies have long time to accumulate. Thus the level of lipofuscin in these cells may be higher than other cells. This may be one of the reasons why lipofuscin bodies are more often seen in nerve tissues and muscular tissues. Differently, in a regenerable tissue such as epidermis, an aged epithelial cell can be soon removed and replaced by a new cell. However, when a tissue is aged, the maintaining efficiency of the tissue will decrease and some aged cells cannot be removed. Long survival of an aged cell enables the accumulation of lipofuscin bodies. Thus, for a regenerable tissue, deposition of a lipofuscin-containing cell is a consequence of aging of the tissue. Deposition of an aged cell is a kind of Misrepair of the tissue. In return, the aged cell can enhance aging of the tissue by affecting the local substance-transportation and information-communication.

An age spot on the skin is composed of basal cells, which are anchoring to the basement membrane in epidermis. Basal cells are the stem cells that are responsible for producing new epithelial cells for regeneration and reparation of epidermis. Basal cells can be injured by UV-radiation and chemical substances, and some of injured cells may survive through Misrepairs and become aged. Normally the aged basal cells, which have reduced functionality, will be soon removed by local langerhans cells and replaced by new basal cells. However, when part of skin has reduced efficiency on maintenance, some of the aged cells cannot be removed. These aged cells can survive for longer time. Therefore, deposition of lipofuscin-containing basal cells in epidermis is a result of aging of the skin.

## 3.2    Enlargement of a flat spot as a result of accumulation of aged cells

A lipofuscin-containing basal cell in skin is actually an invisible age spot, and focalized accumulation of lipofuscin-containing cells results in formation of a visible spot. In our view, a triggering factor for this focalized accumulation of aged cells is the deposition of an aged cell in the tissue. Remaining of an aged cell is a kind of Misrepair, and this Misrepair has altered the relationship of this cell with its neighbor cells. This alteration will affect the functionality of neighbor cells and the local tissue. Thus, an aged cell has two effects on a tissue: **A.** reduced efficiency of neighbor cells on making adaptive responses to environment changes and increased fragility to damage; and **B.** reduced repair-efficiency of local tissue. Thus, the neighbor cells of an aged cell have increased risk for injuries and for Misrepairs. By this mechanism, an aged cell triggers the aging of its neighbor cells.

In addition, because of reduced functionality of local tissue, these aged neighbor cells cannot be completely removed. Thus, they can survive for longer time and make more neighbor cells aging (Figure 2). By such a viscous circle, the range of affected cells is gradually enlarged, and the increase of number of aged cells is accelerated with time. Lipofuscin bodies will accumulate in these long-lived aged cells, and more and more neighbor cells become lipofuscin-containing cells. Accumulation of these cells in a neighborhood results in development of a visible spot that has an irregular shape. In this stage, the spot is flat, since the aged cells are in a normal cell organization in epidermis. Increase of number of affected basal cells results in flat growing of a spot. A spot can rest flat for many years. Only when some of the aged cells die and release lipofuscin bodies, the spot will become protruding.

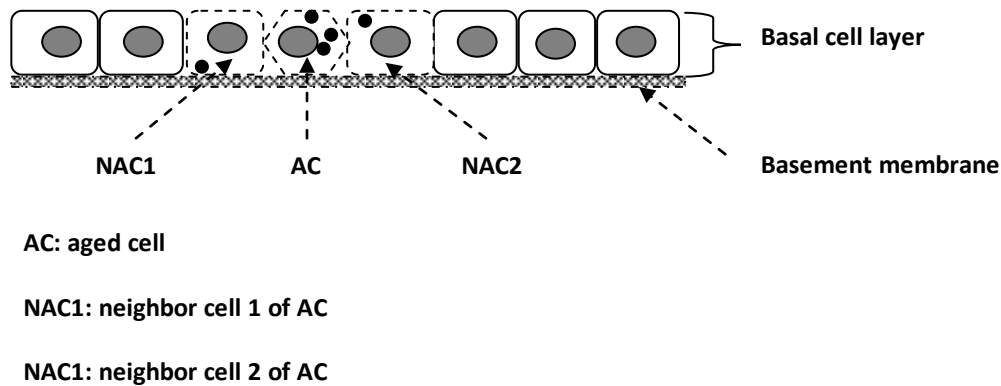

**Figure 2. Effect of an aged cell on a tissue: triggering aging of its neighbor cells**

An aged cell in a tissue will affect the functionality of its neighbor cells, and makes the neighbor cells and the local tissue have increased damage-sensitivity and reduced repair-efficiency. The neighbor cells then have increased risk for injuries and for Misrepairs. In this way, an aged cell (**AC**) triggers aging of its neighbor cells (**NAC1** and **NAC2**).

### 3.3 Protruding of a spot as a result of isolation of lipofuscin bodies in a tissue

On the skin of old people aged over 70 years old, the age spots can change enormously within one year: a spot or part of a spot can become protruding and darker in color. A protruding spot has some characteristics, which can reveal something. **A.** A protruding spot is soft and full of lipids, implying death of lipofuscin-containing cells and release of lipofuscin bodies from dead cells; **B.** A protruding spot cannot drop off from the skin, implying that the spot is probably fixed to basement membrane or to derma; **C.** In a protruding spot, the basal cells are arranged in a column starting from basement membrane to skin surface, implying that the spot is probably fixed to basement membrane. By analyzing these characteristics, we hypothesize that a protruding spot develops as a result of death of lipofuscin-containing cells and isolation of lipofuscin bodies by a capsule. Briefly, a protruding spot may develop by three steps: **1).** death of a lipofuscin-containing cell in flat spot and release of lipid bodies; **2).** isolation of lipofuscin bodies and dead cells *in situ* by constructing a capsule made of fibrotic membrane; and **3).** successive deaths of local lipofuscin-containing cells and enlargement of the capsule by including more and more dead substances, appearing as "growing" of a spot in three-dimension.

An aged cell will die from failure of functionality. The released lipofuscin bodies from dead cells will promote tissue response to remove them. However, lipofuscin bodies are difficult to be digested by langerhans cells, and a solution is to isolate them *in situ*. Isolation of un-degradable substances by a fibrotic capsule is a way of repair/maintenance of a tissue, and it is a Misrepair. Such a Misrepair does not only reduce the toxicity of dead substances to local cells but also rebuilds the structural integrity of the tissue. Basal cells are normally anchored to basement membrane. When a basal cell dies, the neighboring basement membrane can be used as the material for isolating dead substances. A fibrotic capsule is then constructed around the dead cell by producing and modeling a membrane similar to basement membrane (Figure 3A and 3B). Since basement membrane is invisible in hematoxylin and eosin staining, the fibrotic capsule may be unobservable.

When more neighbor cells die, the enwrapping will take place in a higher level with a new layer of fibrotic membrane. The new membrane will enwrap both of the newly dead substances and the old capsule. Successive deaths of cells make the capsule include more and more lipofuscin bodies with more and more layers of membrane. In this way, the capsule becomes bigger and bigger in three-dimension and the spot becomes protruding. Concentrated lipofuscin bodies in the capsule make the protruding spot soft and have deeper color. Multiple layers of fibrotic membranes in the capsule make the spot look to be covered by cuticle-like substances. Since the fibrotic membranes in capsule are fixed to basement membrane, the spot is irremovable.

Our hypothesis on the process of development of a protruding spot is schematically presented in Figure 3. At first, death of a lipofuscin-containing cell promotes development of the first-level of capsule made of a layer of fibrotic membrane (**Capsule 1,** Figure 3B). Reorganization of local basal cells will take place for sealing the epidermis. Then, death of neighbor cells next to the first one promotes development of the second-level of capsule by another layer of fibrotic membrane (**Capsule 2,** Figure 3C), which enwraps the newly dead cells and capsule 1. In this way, the capsule becomes bigger and bigger by including old capsules (capsule 1 and 2) and new dead substances (**Capsule 3,** Figure 3D). Successive enwrapping of dead cells and lipofuscin bodies in more and more layers of fibrotic membranes makes part of a spot "grow" in 3D.

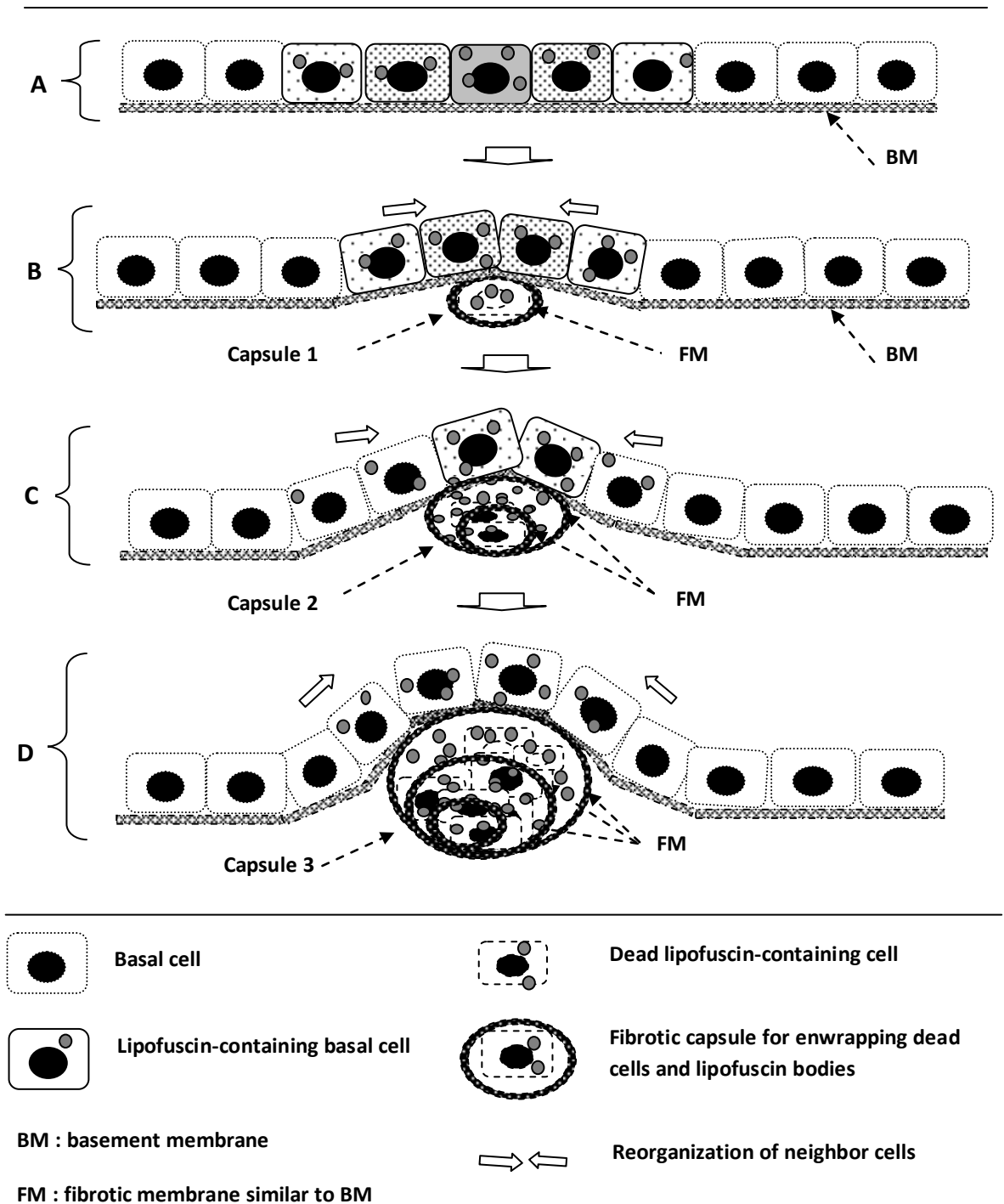

**Figure 3. A hypothesized process of development of a protruding spot**

Our hypothesis on the process of development of a protruding spot is schematically presented. At first, death of a lipofuscin-containing cell promotes development of the first-level of capsule made of a layer of fibrotic membrane (**Capsule 1**, **A** and **B**). Reorganization of local basal cells will take place for sealing the epidermis (**B, C,** and **D**). Then, death of neighbor cells next to the first one promotes development of the second-level of capsule by another layer of fibrotic membrane (**Capsule 2**, **C**), which enwraps the newly dead cells and capsule 1. In this way, the capsule becomes bigger and bigger, by including old capsules (capsule 1 and capsule 2) and new dead substances (**Capsule 3, D**). Successive enwrapping of dead cells and lipofuscin bodies in multiple levels of capsules by multiple layers of fibrotic membranes makes part of a spot protruding.

## IV.    Conclusions

Aging of a tissue is the basis for development of age spots. Development of an age spot proceeds in three steps: **A**. deposition of a lipofuscin-containing cell in an aged tissue, which determines the location of a spot; **B.** accumulation of aged cells in neighborhood, which results in development and enlargement of a flat spot; and **C.** death of lipofuscin-containing cells and isolation of lipofuscin bodies by a fibrotic capsule, which results in protruding of part of a flat spot in the skin. Thus, development of an age spot is a result of accumulation of aged cells in aged skin. The in-homogeneous distribution of age spots is a result of focalized accumulation of aged cells.